\documentclass[aps,prl,twocolumn,amsfonts,showpacs,longbibliography,citeautoscript,superscriptaddress,nofootinbib]{revtex4-2}
\usepackage[toc,page]{appendix}
\usepackage[british]{babel}
\usepackage[utf8]{inputenc}
\usepackage{amsmath,amssymb}
\usepackage{graphicx}
\usepackage{ae}
\usepackage{esdiff}
\usepackage{braket}
\usepackage{bbm}
\usepackage{simplewick}
\usepackage{float}
\usepackage{braket}
\usepackage{simplewick}
\usepackage{color}
\usepackage{framed}
\usepackage[caption=false]{subfig}
\usepackage[pdftex,colorlinks=true,linkcolor=blue,citecolor=blue,urlcolor=black]{hyperref}
\usepackage{hyperref}
\usepackage{mathtools}
\usepackage{enumitem}
\usepackage{ragged2e}
\usepackage{floatflt}
\usepackage{import}
\usepackage{titlesec}
\usepackage{multirow}
\usepackage{etoolbox}
\usepackage{appendix}
\usepackage{dsfont}
\usepackage{comment}
\usepackage{xcolor}
\usepackage{blindtext}
\usepackage[makeroom]{cancel}
\usepackage{mathtools}

\renewcommand{\vec}[1]{\boldsymbol{#1}}

\begin{document}

\newcommand{\norm}[1]{\left\lVert#1\right\rVert}

\def \r{{\boldsymbol{r}}}
\def \k{{\boldsymbol{k}}}
\def \p{{\boldsymbol{p}}}
\def \q{{\boldsymbol{q}}}
\def \m{\boldsymbol{m}}
\def \h{\boldsymbol{h}}
\def \e {\boldsymbol{e}}
\def \x{{\textbf{x}}}
\def \A{{\textbf{{A}}}}
\def \a{{\textbf{{a}}}}
\def \b{{\textbf{{b}}}}
\def \c{{\textbf{{c}}}}
\def \z{{\textbf{{z}}}}
\def \0{{\boldsymbol{{0}}}}
\def \dl{\frac{\partial}{\partial l}}
\def \P{{\boldsymbol{P}}}
\def \K{{\boldsymbol{K}}}
\def \sigmad{{\sigma_{\downarrow\uparrow}}}
\def \uone{{\boldsymbol{u}_1}}
\def \utwo{{\boldsymbol{u}_2}}
\def \edown{{\epsilon_{\downarrow}}}
\def \omfl{{\omega_{\text{FL}}}}
\def \piph{\Pi_\text{ph}}
\def \sign{ \text{sign}}
\def \lamt{\tilde{\lambda}}
\def \Sigmab{\Sigma_{\omega^2, \text{bs}}}
\def \intk{{\int_\textbf{k}}}
\def \Ims{\text{Im} [ \Sigma^R(\omega) ]}
\def \Gammabs{\Gamma_{\text{bs}}}
\def \aq{|\q|}
\def \ak{|\k|}
\def \omt{\omega_{\text{th}}}
\def \beq{\begin{eqnarray}}
\def \eeq{\end{eqnarray}}
\def \nn{\nonumber}

\definecolor{mgrey}{RGB}{63,63,63}
\definecolor{mred}{RGB}{235,97,51}
\newcommand{\mg}[1]{{\color{mgrey}{#1}}}
\newcommand{\mr}[1]{{\color{mred}{#1}}}

\newcommand{\red}[1]{{\color{red}{#1}}}
\newcommand{\blue}[1]{{\color{blue}{#1}}}

\newcommand{\DC}[1]{ { \color{red} 
\footnotesize (\textsf{DC})\textsf{\textsl{#1}}}}

\newcommand{\DP}[1]{ { \color{blue} 
\footnotesize (\textsf{DP})\textsf{\textsl{#1}}}}

\def\BigColSep{\setlength{\arraycolsep}{50pt}}

\title{Dynamical Control of Superconductivity in Superconductor–Ferromagnet Bilayers}
\author{Dimitri Pimenov}\email{dp589@cornell.edu}
\affiliation{Department of Physics, Cornell University, Ithaca, New York 14853, USA}
\author{Nicholas R. Poniatowski}
\affiliation{Department of Physics, Harvard University, Cambridge, Massachusetts 02138, USA}
\author{Charlotte G.~L. B\o ttcher}
\affiliation{Department of Applied Physics, Stanford University, Stanford, California 94305, USA}
\author{Amir Yacoby}
\affiliation{Department of Physics, Harvard University, Cambridge, Massachusetts 02138, USA}
\author{Debanjan Chowdhury} \email{debanjanchowdhury@cornell.edu}
\affiliation{Department of Physics, Cornell University, Ithaca, New York 14853, USA}
\date{\today}

\begin{abstract}
We study a simplified model of a ferromagnetic metal proximitized by a fully-gapped $s-$wave superconductor and integrated with a microwave resonator. The low-energy excitations in the combined system consist of ferromagnetic magnons, Bogoliubov excitations of the superconductor, and cavity photons. We show here that when the magnons and photons have comparable frequencies and are subject to an external drive, the hybridized driven magnon-polaritons induce a non-equilibrium crossover from the expected proximitized nodal $p-$wave superconductor to a fully gapped $(p_x+ip_y)-$superconductor. Moreover, the characteristic crossover temperature is inversely related to the magnon-photon detuning. We compute the temperature-dependent renormalization of the cavity photon frequencies across this nodal to nodeless evolution, which modifies the kinetic inductance of the resonator, and find a number of non-trivial features tied to the non-equilibrium (i.e., driven) nature of the problem. We compare and contrast these results with a recent circuit quantum electrodynamics (cQED) based experiment studying a permalloy-niobium bilayer, where a non-trivial dependence of the low-temperature cavity response on the magnon-photon detuning was observed. Our results pave the way for a principled exploration of engineering novel states of matter by coupling cavity photons to electronic collective modes in correlated two-dimensional materials and interfaces. 
\end{abstract}
\maketitle

\textit{Introduction.-} The spectroscopy of micron-scale materials, including moir\'e heterostructures, has long posed an experimental challenge due to the simultaneous requirements of high-energy resolution and strong coupling to small-volume samples.
On-chip microwave resonators, naturally described within the framework of circuit quantum electrodynamics \cite{RevModPhys.93.025005} (cQED), are a promising platform to meet both these needs.  In a typical cavity setup, the resonator is coupled to an external transmission line, which can be used to probe the resonator response at its resonant frequency $\omega_r$, while simultaneously enabling an external microwave drive. When $\omega_r$ is resonant with collective modes in the heterostructure, this drive can probe these excitations, as well as dynamically alter its many-body state and modify its excitation spectrum --- furnishing a promising platform for cavity control of quantum materials \cite{RevModPhys.93.041002, Mivehvar02012021, 10.1063/5.0083825, PhysRevLett.104.106402, Goldie_2013,   PhysRevLett.112.047004}. 

Recent studies have reported measurements of the diamagnetic response of two-dimensional mesoscopic superconductors using variants of this technique 
 \cite{PhysRevLett.128.107701, magnonSCpaper,  10.21468/SciPostPhys.16.5.115, PhysRevResearch.6.043245, PhysRevApplied.22.034004,TanakaTBG, BanerjeeTTG, JinvDW, zaman2025kineticinductancefewlayernbse2, chistolini2025contactlesscavitysensingsuperfluid}. These measurements have focused on the superfluid stiffness, $\rho_s(T)$, which characterizes the robustness of the superconductor with respect to phase fluctuations. In the simplest model of the resonator coupled to the low-energy (Bogoliubov) excitations in the superconductor, the diamagnetic, inductive response of the superconductor alters the  microwave photon frequency. The temperature-dependence of $\rho_s(T)$ allows one to distinguish between nodal and nodeless superconductors, and can be extracted from the variation in the resonator frequency,
$\delta\omega_r(T)$, as a function of temperature. This is analogous to the standard determination of the gap symmetry from the temperature-dependence of the penetration depth for bulk samples \cite{PhysRevLett.70.3999, prozorov2006magnetic}.

\begin{figure}
\centering
\includegraphics[width= \columnwidth]{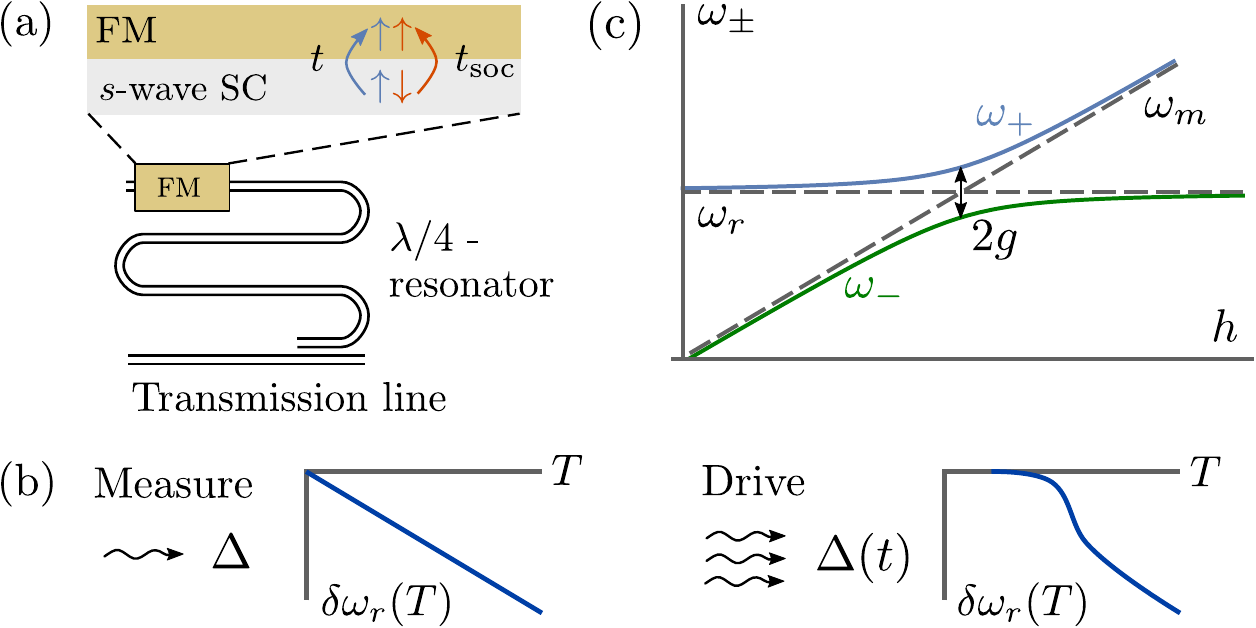}\caption{(a) A superconductor (SC) - ferromagnet (FM) bilayer integrated with a microwave ($\lambda/4$-) resonator, accessed via a transmission line. The normal ($t$) and spin-orbit assisted ($t_{\rm{soc}}$) tunneling induces triplet superconductivity in the FM layer; see Eq.\ \eqref{Hamiltonian}. (b) Left: Within linear-response, temperature-dependent shift of the resonator frequency $\delta \omega_r(T)$ probes the equilibrium nodal-gap $\Delta_\k$ of the superconductor. Right: Driving renders $\Delta(t)$ time-dependent and nodeless, modifying $\delta \omega_r(T)$. (c) Magnon-polariton dispersion in the vicinity of a resonance where the resonator frequency $\omega_r$ is comparable to the magnon frequency $\omega_m$, where the latter is tunable with an external magnetic field, $h$. }
\label{main_fig}
\end{figure}

Motivated by a recent experiment carried out by some of the present authors \cite{magnonSCpaper}, in this letter we revisit the driven response of a nodal superconductor coupled to a resonator, which appears at the interface between a gapped superconductor and a metallic ferromagnet (FM) [see Fig.\ \ref{main_fig}(a)]. We investigate the characteristic feedback of the driven superconductor on $\delta\omega_r(T)$  [Fig.\ \ref{main_fig}(b)] within a simplified model of the microwave photons hybridizing with the magnons in the FM layer in the limit of a weak, slow drive at low temperatures. We find a number of new results, which includes a possible resolution to the previously unexplained appearance of a sharp upturn in $\delta \omega_r(T)$ at low temperature that manifested only for small magnon-photon detunings, when the hybrid resonator response is measured near the ferromagnetic resonance. Our theoretical analysis proceeds in three key steps: (i) the driven photon-magnon dynamics, (ii) their impact on the superconducting state, and (iii) the resulting lineshapes of $\delta\omega_r(T)$.

\textit{Driven Magnon-Polaritons.-} 
We consider a superconducting microwave resonator hosting long-lived photonic modes with energy $\omega_r$, working in units where $\hbar = c = \mu_0 = 1$. A metallic ferromagnetic layer is placed in direct electrical contact with the superconducting resonator with a conventional $s-$wave pairing. The magnetization of the ferromagnet $\vec{m}$ is controlled by an external in-plane field $\vec{h}_0$, specifically $\vec{h}_0 = h_x \hat{e}_x$ and $\vec{m} = m_x\hat{e}_x$ (with $\hat{e}_z$ pointing out of plane). The elementary excitation of the ferromagnet is the ferromagnetic resonance, the uniform collective precession of the magnetization. It occurs at frequency $\omega_m$, as can be described classically by the Landau-Lifshitz-Gilbert equation~\cite{stancil2009spin,PhysRevLett.99.057003,boettcher2022new}. Microscopically, the magnetization can be related to a macro-spin $\vec{S}$ via $\vec{m} = \mu_e \vec{S}/V$, where $V$ is the ferromagnet's volume and $\mu_e$ is the electron magnetic moment. Quantizing $\vec{S}$ fluctuations through a leading-order Holstein-Primakoff expansion yields Kittel magnons --- the quantum analog of ferromagnetic resonance --- with the same characteristic frequency $\omega_m$. In two dimensions, the magnon frequency takes the form, $\omega_m \simeq \gamma\sqrt{(m_x +h_x)h_x}$,
where $\gamma$ is the gyromagnetic ratio. Throughout our analysis, we will treat $\omega_m$ as a tunable free parameter, controllable via the external magnetic field $h_x$.

In the rotating wave approximation, the coupled photon-magnon system is then described by the Hamiltonian 
\beq
\label{Hmp}
H_\text{mp} =  \omega_r~\hat{a}^\dagger \hat{a}  + \omega_m~\hat{b}^\dagger \hat{b}  + i g(\hat{a}^\dagger \hat{b} - \hat{b}^\dagger \hat{a}) \ ,  
\eeq
where $g$ is the magnon-photon coupling. Here $a,~a^\dagger$ and $b,~b^\dagger$ are the bosonic annihilation and creation operators associated with the resonator and magnon modes, respectively.  The eigenmodes of Eq.\ \eqref{Hmp} take the form 
\beq
\label{mpequation}
\omega_\pm = \frac{1}{2} \left(\omega_r + \omega_m\pm \sqrt{(\omega_r - \omega_m)^2 + 4g^2}\right) \ ,
\eeq      
representing the two magnon-polariton branches \cite{haug2009quantum, PhysRevLett.125.237201,  PhysRevApplied.20.024039, PhysRevApplied.22.034004}; see Fig.\ \ref{main_fig}(c).

Let us now switch on an external monochromatic drive for the photons, $H_d(t) = V_d \exp(-i\omega_d t) a^\dagger$ + h.c., which is assumed to be near-resonant with the upper polariton branch $\omega_+$. This drive leads to a precessing magnetization. We focus on the out-of-plane component $m_z$, which has the strongest influence on the induced SC state, as will be shown below. Assuming $m_z \ll m_x$, it can be defined as $m_z = m_x/\sqrt{2S} \braket{b + b^\dagger}$, and takes the form 
\begin{subequations}
\beq
\label{mzmain}
m_z(t) &=&  A(\omega_r,\omega_m)\sin(\omega_d t) \frac{V_d m_x/\sqrt{S} }{\sqrt{(\omega_d - \omega_+)^2+ (\kappa  \omega_+)^2}},\nn\\ \\
A(\omega_r,\omega_m) &=& \frac{\sqrt{2}g }{\sqrt{4g^2 + (\omega_r - \omega_m)^2+ (\kappa  \omega_+)^2} }, 
\eeq
\end{subequations}
where $A(\omega_r,\omega_m)$ is determined by a hybridization amplitude between the photon and magnon modes that is independent of the drive frequency, and $\kappa$ is an effective damping constant, which is set by the inverse cavity- and magnon lifetimes, respectively.  The key is to note that the second factor in Eq.\ \eqref{mzmain} shows that the $m_z$ oscillation amplitude is inversely proportional to the proximity between the driving frequency $\omega_d$ and the collective mode frequency $\omega_+$. A purely semiclassical description gives a similar result (see the Supplementary material \cite{suppl}). \\
\textit{Effective FM superconductor.-}
We now turn to a  microscopic description of the superconductor-ferromagnet system, where we incorporate both spin-conserving tunneling as well as spin-flip tunneling due to the spin-orbit coupling at the interface [Fig.~\ref{main_fig}(a)] \cite{PhysRevB.86.054521, PhysRevB.89.134517, RevModPhys.96.021003, magnonSCpaper}. A minimal Bogoliubov-de-Gennes (BdG) Hamiltonian is given by \cite{magnonSCpaper}, 
\begin{subequations}
\label{Hamiltonian}
\beq
{H}_\text{BdG} &=& \begin{pmatrix}
  {H}^{(\text{s})} & {H_\text{tun}} \\ {H_\text{tun}} & {H}^{(\text{f})} + \delta {\hat m} 
  \end{pmatrix}, \\ 
H^{(\text{s})} &=& 
\xi_\k^{(\text{s})} \tau_3 + \Delta \tau_2 \sigma_2, ~~H^{\text{(f)}} =
 (\xi_\k^{(\text{f})} + \mu_e m_x \sigma_1 ) \tau_3,\nn \\ \\ 
 H_\text{tun} &=& t \tau_3 + t_\text{soc}(k_x \tau_3 \sigma_2 - k_y \sigma_1), \\
 \delta \hat m &=& \mu_e (m_y \sigma_2 + m_z \sigma_3)\tau_3, \label{delm}
\eeq   
\end{subequations}
where the Pauli matrices $\tau_i,\sigma_i$ act in Nambu and spin space, respectively. Here, $\xi_\k^{(\text{s})},~\xi_\k^{(\text{f})}$ represent the electronic dispersion in the superconducting and ferromagnetic layers, respectively, and $\Delta$ is the bulk gap in the superconducting layer. We have included regular and spin-orbit-assisted tunnelings, $t$ and $t_{\text{soc}}$, between the two layers. For simplicity, here we view the magnetization in the ferromagnetic layer, $m_x$, as an effective in-plane magnetic field.  As we demonstrate below, the drive-induced magnetization fluctuations, represented by the terms combined under $\delta\hat{m}$ in Eq.~\eqref{delm}, and ignored in previous treatments, lead to qualitatively new effects.

In what follows, we take $m_x \gg m_y, m_z$, and derive an effective description for the {\it majority} spins in the ferromagnet, integrating out the minority spins and the superconducting degrees of freedom (see the Supplementary material \cite{suppl}). 

For vanishing external frequencies and to leading order in $t_\text{soc}$, we obtain the following self-energy for the majority spins \cite{suppl}: 
\begin{subequations}
\beq
\label{Sigmadown}
\Sigma(\omega = 0, \k)  &\equiv& \Sigma_1 + \Sigma_2 + \Sigma_3, \\  
\Sigma_1(\omega=0,\k) &=&  \frac{1}{2}\frac{ t^2 \delta\xi_\k}{(\delta \xi_\k)^2 + \Delta^2} \tau_3 +  k_y  \frac{t t_\text{soc} \delta\xi_\k}{(\delta \xi_\k)^2 + \Delta^2} \tau_0,  \quad \quad  \\
\Sigma_2(\omega=0,\k) &=& k_x  \frac{t t_\text{soc} \Delta}{(\delta \xi_\k)^2 + \Delta^2} \tau_2, \\
\Sigma_3(\omega=0,\k) &=& k_y \frac{tt_\text{soc}\Delta}{(\delta \xi_\k)^2 + \Delta^2} \cdot \frac{2\mu_e m_z}{\mu_e m_x + \xi_\k^{(\text{f})}} \tau_1 , 
\eeq
\end{subequations}
where $\delta \xi_\k = \xi_\k^{(\text{s})} - \xi_\k^{(\text{f})}$.
The normal contribution $\Sigma_1$ represents a renormalization of the dispersion, while $\Sigma_2$ is the proximitized nodal spin-triplet component; both terms have been discussed previously \cite{magnonSCpaper}. The effects of the new contribution, $\Sigma_3$, have not been explored so far. 
This new term shows that the out-of-plane canting, $m_z$, can lead to $p_x + ip_y$ pairing. While this has previously been derived for a static out-of-plane field in Ref.\ \cite{PhysRevB.86.054521}, here we explore the effects of the dynamically induced $m_z$-fluctuations on the pairing state. 

\begin{figure}
\centering
\includegraphics[width= \columnwidth]{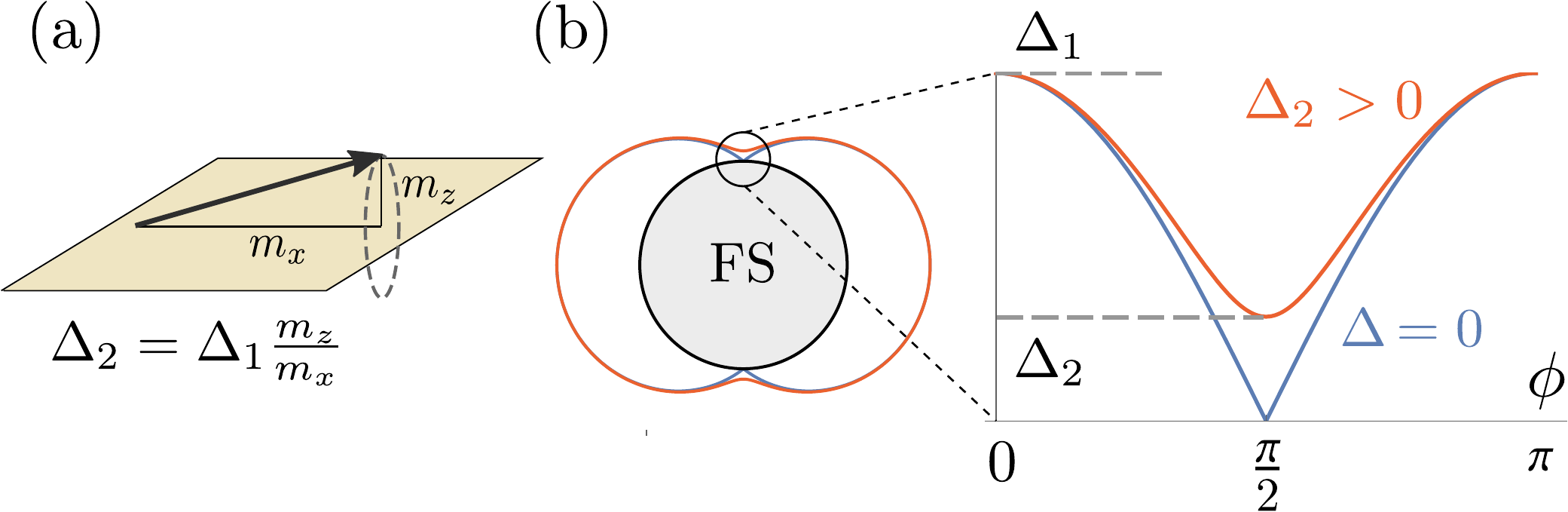}
\caption{(a) Secondary gap component induced by  an out-of-plane canting of the magnetization. (b) Sketch of the induced gap, Eq.\ \eqref{Deltak},  with $\phi$ the angular direction along the Fermi surface.}
\label{gapsketch}
\end{figure}

In the vicinity of the majority spin Fermi level, $\xi_\k^{(\text{f})} \simeq \mu_e m_x$, and ignoring $\Sigma_1$, the leading contribution to the self-energy can then be expressed as
\begin{subequations}
\begin{align}
\label{Sigmadownind}
&\Sigma(\omega = 0, \k) \approx \hat{k}_x \Delta_1 \tau_2 + \hat{k}_y \Delta_2 \tau_1    =   
 \begin{pmatrix} 0 & -i \Delta_\k  \\ 
i \Delta_\k^*& 0  \end{pmatrix},  \\ &    \label{Deltak}
 \text{where}  \ \hat{k}_i = \frac{k_i}{k_F},\ 
   \Delta_\k = \hat{k}_x \Delta_1 + i \hat{k}_y \Delta_2, \\ &  \text{and} \    \Delta_1= \frac{\Delta t t_{\text{soc}}k_F}{[(\mu_e m_x)^2 +  \Delta^2]}, \ \Delta_2 = \frac{m_z}{m_x} \Delta_1.
   \label{Delta2Delta1}
\end{align}
\end{subequations}
A sketch of the resulting gap is shown in Fig.\ \ref{gapsketch}. For a drive that is slow enough compared to fermionic scales ($\omega_d \rightarrow 0)$, an effective instantaneous gap $\Delta_\k(t)$ can be obtained by inserting the  $m_z(t)$ into Eq.\ \eqref{Delta2Delta1}.

{\it Equilibrium diamagnetic response.-}
Let us now recap how the superconducting state can be observed experimentally, starting from the equilibrium situation for a vanishing drive. The vector potential, $\vec{\hat{A}}$, associated with the photons couples to mobile electrons in the ferromagnetic layer via the usual para- and diamagnetic terms,
$
\vec{\hat{J}}\vec{\hat{A}} + \frac{e^2}{2m} \hat{n} (\vec{\hat{{A}}})^2$, where $ 
\vec{\hat{A}}  =  \sqrt{\frac{{2\pi}}{\omega_r}} \vec{u} (a + a^\dagger)
$, 
and $\vec{u}$ is the polarization vector. $\vec{\hat{J}}, \hat{n}$ represent the fermionic current and density operators, with $e$ and $m$ the electronic charge and mass, respectively. For simplicity, we have assumed a Galilean invariant form of the electron dispersion.

In the superconducting state, the coupling of the low energy degrees of freedom to the cavity photons leads to a temperature-dependent correction of their frequency via the  Anderson-Higgs mechanism, $\delta H_{\rm{eff},a} = \lambda_a a^\dagger a~ {\rho_{s} (T)}$, where $\lambda_a$ is a dimensionless constant. Here, $\rho_s$ is the superfluid stiffness evaluated via the conventional transverse electromagnetic response. This leads to a measurable temperature-dependent  shift of the cavity frequency $ \delta \omega_r(T) \sim \rho_{s}(T) - \rho_{s}(0)$, whose $T$-scaling  reflects the gap structure and low-lying excitation spectrum in the superconductor. 
 Given the form of the hybridization in Eq.\ \eqref{mpequation}, the $T$-dependence from $\delta \omega_r(T)$ is also inherited by the magnon-polariton modes, as also recently observed in Refs.\ \cite{PhysRevApplied.22.034004, ghirri2025coherentcouplingybcosuperconducting}. 
 In this polaritonic regime, the temperature-dependence of the superfluid stiffness can then be extracted from the shift in the polariton frequencies.

Quantitatively, the frequency shift $\delta \omega_r(T)$
can be extracted from the correlation function \begin{subequations}
\beq
\label{Pifirst}
\Pi(\omega_r, T) &=& -\braket{ \hat{J}_{\vec{u}}(i\omega) \hat{J}_{\vec{u}}(-i\omega) }\bigg|_{i\omega \rightarrow \omega_r + i0^+}, \\
\delta \omega_r(T) &\propto& -\left( \text{Re}\left[\Pi(\omega_r, T)\right]  - \text{Re}\left[\Pi(\omega_r, 0)\right] \right),\label{deltaPiom}
\eeq
\end{subequations}
where $\hat{J}_{\vec{u}}$ is the component of the  current operator along the field direction $\vec{u}$, and the expectation value $\braket{...}$ is evaluated with respect to the thermal many-body state.
Since the characteristic wavelength associated with the microwave probe is much larger than the typical sample size, it is justified to set the external momentum $q\rightarrow0$ in the above response function; this is consistent with treating the cQED setup as a lumped-element circuit \cite{RevModPhys.93.025005} and attributing the frequency shift $\delta \omega_r(T)$ to the kinetic inductance of the superconductor.

The response function is evaluated at the frequency $\omega_r$ of the  photons (or magnon-polaritons near a resonance). When $\omega_r \rightarrow 0$, the correlation function is equivalent to paramagnetic contribution to the change in the superfluid stiffness. 
In the clean limit, Eq.\ \eqref{deltaPiom} evaluates to \cite{suppl}
\begin{align} 
\label{RePi}
\text{Re}\left[\Pi(\omega_r , T)\right] =   \frac{1}{2m^2}  \int_\k k_{\vec{u}}^2   \frac{\sinh{\frac{\omega_r}{T}}}{\omega_r \left(\cosh{\frac{E_\k}{T}} + \cosh{\frac{\omega_r}{T}}\right)} \ , 
\end{align} 
where the $\sim k_{\vec{u}}^2$ contribution arises from the paramagnetic current vertex evaluated on the Fermi surface (with $|k_{\boldsymbol{u}}| \sim k_F$), and $E_\k=\sqrt{\xi_\k^2+|\Delta_\k|^2}$. 

We can recover some of the standard limits from the above expression as follows. For a fully gapped spectrum with $E_\k=\Delta$ at $k=k_F$ and $\omega_r \rightarrow 0$, we obtain $\delta\omega_r(T) \sim - \exp(-\Delta/T)$. For a nodal $p-$wave order-parameter, the frequency shift depends on the relative orientation of the microwave current and the nodal direction \cite{magnonSCpaper}: When $\boldsymbol{A} \sim \vec{u}$ points along the nodal direction, one obtains $T$-linear behavior $\delta\omega_r(T) \sim -T$, while for a perpendicular orientation, one finds $\delta\omega_r(T) \sim -T^3$. The difference in the behavior, which was indeed observed in \cite{magnonSCpaper}, is due to the distinct decay channels for the microwave photons into the low-energy Bogoliubov quasiparticles. 

For large detuning between cavity photon and magnon modes, $\omega_r - \omega_m \gg g$, the experiment \cite{magnonSCpaper} indeed observed a frequency shift $\delta\omega_r(T)$ consistent with the above power laws (with corrections due to disorder), supporting the existence of an equilibrium proximity-induced $p-$wave superconductor. However, significant deviations from these power laws were also observed in the polaritonic regime, $\omega_r \rightarrow \omega_m$, at asymptotically low temperatures. To develop a better understanding for the possible origin of these effects, we now study the impact of an induced time-dependent gap $\Delta_\k(t)$ as given by Eq.\ \eqref{Deltak} with a finite out-of-plane magnetization $m_z(t)$. 

{\it Driven diamagnetic response.-} 
Instead of solving the full time-dependent problem, we estimate $\delta\omega_r(T)$ by appealing to an ``adiabatic" approximation: we assume that the driving frequency $\omega_d$ is the smallest energy scale in the problem, and the fermionic time scales are fast compared to the slow drive. In the experiment \cite{magnonSCpaper}, $\omega_d/2\pi  \simeq 5 \text{GHz} \approx 0.25$K is small compared to $\Delta_1\gtrsim 2$K.\footnote{The energy scale $\Delta_1$ was not obtained directly from the experiment \cite{magnonSCpaper}, but the linear nodal behavior of $\delta \omega_r$ associated with $T \ll \Delta_1$ persists in the full measured range $T < 0.8$K.} In this adiabatic limit,  we can evaluate $\text{Re}[\Pi(\omega_r, T)]$ from Eq.\ \eqref{RePi}  instantaneously for a given $\Delta_\k(t)$, which enters through $E_\k =  \sqrt{\xi_\k^2 + |\Delta_\k(t)|^2}$. In the experiment, the external drive
is simultaneously used to drive the system and measure the cavity frequency $\omega_r$. To achieve a frequency resolution which is small compared to $\omega_r$, the drive needs to be switched on for a long time $ \gg \omega_d^{-1}$. Therefore, the effective energy shift $\delta \omega_r$ is given by the period-averaged expression, 
\begin{align} 
\label{timeaverage}
\overline{\text{Re}\left[\Pi(\omega_r , T)\right] } = \frac{1}{T_\text{drive}} \int_0^{T_\text{drive}} dt~{\text{Re}\left[\Pi(\omega_r , T)\right] }, 
\end{align}
where $T_\text{drive} = 2\pi/\omega_d$ and the only explicit time dependence in the integrand appears through $\Delta_2(t)$. We parametrize this time dependence as  $\Delta_2(t) = \Delta^*\sin(\omega_d t)$, with $\Delta^*$ determined by Eqs.\ \eqref{Delta2Delta1}, \eqref{mzmain}. Since $\omega_r \approx \omega_d$ is a small scale, we focus on the evaluation of 
$\delta \omega_r(T) \equiv \overline{\text{Re}[\Pi(\omega_r \rightarrow 0,T)]}$ 
 in the limit $\omega_r \rightarrow 0$; the role of finite frequencies is briefly discussed in the Supplementary material \cite{suppl}.

\begin{figure}
\centering
\includegraphics[width= \columnwidth]{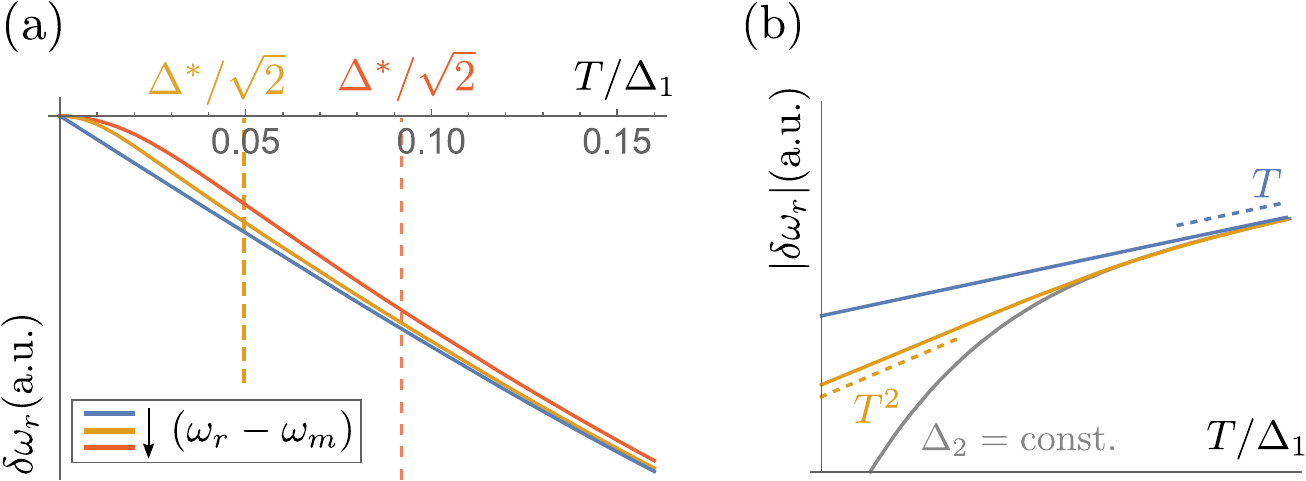}
\caption{Cavity frequency shift $\delta \omega_r$ determined from Eq.\ \eqref{timeaverage}: 
(a) ``upturn" due to a temperature-independent oscillatory secondary gap $\Delta_2(t) = \Delta^*\sin(\omega_d t) $ for decreasing values of $(\omega_r - \omega_m)$; the dashed vertical lines show the corresponding time-averages $\Delta^*/\sqrt{2}$.   (b) Log-Log plot of $|\delta \omega_r|$. The gray curve shows $\delta\omega_r$ evaluated with a time-independent  gap $\Delta_2 = 0.5\Delta^*$  for comparison. }
\label{RePifigs}
\end{figure}

Our naive expectation for a nodeless superconductor is an exponential suppression of $\delta\omega_r(T)$ at low temperatures. However, when evaluating $\delta \omega_r(T)$ for a time-dependent $\Delta_2(t)$, note that there are certain time intervals over which $T>\Delta_2(t)$, while $T<\Delta_2(t)$ over the remaining time interval. For the former, the  response should not reflect that of a fully gapped superconductor. Thus, when computing the period averaged response, the ``weighted" frequency shift will not generically correspond to that of a fully gapped superconductor at any finite temperature. 
We can therefore expect the following behavior in two asymptotic limits:  For $T \gg \Delta^*$, the low-energy gap is not important, and  $\delta \omega_r(T)$ approaches its $\Delta_2 = 0$ value. For $T \ll \Delta^*$,  $\delta \omega_r(T)$ is dominated by times $t$ such that $\Delta_2(t) < T$, which give a $T$-linear or quadratic contribution, depending on the relative orientation of $\vec{A}$ and the nodal direction. The window of times with $\Delta_2(t) < T$ is itself of order $T$, which leads to a suppression of $\delta \omega_r(T)$ by an additional power of $T$ compared to the equilibrium result.

{\it Low-temperature crossover.-} A typical plot of $\delta \omega_r(T)$ obtained from the numerical integration in Eq.~\eqref{timeaverage} for different values of $\omega_r - \omega_m \approx \omega_d-\omega_+$ is shown  in Fig.\ \ref{RePifigs}(a) for $\vec{A}$ oriented in the nodal direction; the perpendicular orientation leads to a qualitatively similar outcome. The absence of any exponential suppression is clearly visible. In Fig.\ \ref{RePifigs}(b), we show the modulus of the same quantity on a log-log plot, verifying the expected crossover behavior. One observes a modest ``upturn" effect below a temperature $T\sim\Delta^*$. The frequency shift is suppressed for $T < \Delta^*$ where the system effectively has a nodeless $p_x + i p_y$ gap  superimposed over the equilibrium nodal gap, reducing the time-averaged density of low lying states.

The recent experiment \cite{magnonSCpaper} has indeed observed signs of an upturn below a low temperature, $T^*$, that is inversely related to the proximity to the resonance, $|\omega_r - \omega_m|$. Interestingly, the anisotropic nodal response discussed previously disappears below $T^*$, suggesting the likely emergence of an isotropic nodeless state. These features are in broad agreement with our theoretical picture discussed above if we associate $T^* \sim \Delta^*$, with a frequency dependence inherited from the amplitude of $m_z(t)$ in Eq.\ \eqref{mzmain}: for $T < \Delta^*$, $\delta \omega_r$ probes the density of states of Bogoliubov quasiparticles at energies $< \Delta^*$ which is drastically reduced compared to the nodal case. It is interesting to note that within our theoretical setup, the asymptotic low-temperature state is always the drive-induced $p_x+ip_y$ state of Eq.~\eqref{Deltak}, but its onset temperature becomes progressively lower with increasing distance away from the resonance, in qualitative agreement with observations in \cite{magnonSCpaper} within the experimentally accessible temperature range, $T \gtrsim 50$ mK.

{\it Outlook.-} We have shown here that microwave drives in the vicinity of the ferromagnetic resonance can alter the nature of the superconducting state associated with a ferromagnet-superconductor bilayer, turning it from a nodal $p$ to a nodeless $(p_x+ip_y)$ state. Unlike previous proposals which have suggested that it can impact the spin structure of the superconducting layer \cite{PhysRevB.96.024414, PhysRevB.104.144428, PhysRevB.80.104416, Morten_2008,  PhysRevLett.99.057003, PhysRevB.107.184437}, our proposal suggests the richer possibility of dynamically engineering new forms of pairing symmetry \cite{PhysRevLett.127.147701}. While we have focused only on the ``adiabatic" limit of the driven problem, it would be interesting to analyze the fully time-dependent solution for the superconducting state, and extend it to other time-dependent scenarios involving quenches \cite{PhysRevLett.93.160401, neri2025collectivemodespectroscopytimereversal}. The effects of heating due to enhanced magnon-electron scattering are also worthy of further analysis \cite{PhysRevB.94.214504, Goldie_2013, PhysRevApplied.19.054087}. Finally, there is a rich set of possibilities that remains to be explored by integrating a resonator with distinct magnetically ordered states and simple superconductors to design novel non-equilibrium superconductors.

There are a number of questions that continue to remain about the experimentally observed ``upturn". While the degree of quantitative upturn certainly depends on the microscopic parameters, it is noticeably larger in the experiment compared to our theory. One possibility is that in the driven system and going beyond the mean-field approximation, the Bogoliubov quasiparticles undergo scattering with a temperature-dependent rate that affects $\delta\omega_r(T)$ \cite{suppl}. It would also be useful to include the effects of thermal fluctuations self-consistently. A finite thermal magnon population associated with $\braket{m_z^2}\sim\braket{b^\dagger b}$ is expected for $T \gtrsim \omega_m$, which also affects $\delta\omega_r(T)$ via the Bogoliubov dispersion, $E_\k$. However, treating the magnons as free quasiparticles, this ``thermal gap"  increases with temperature, which is not sufficient to account for the upturn.

On the experimental side, it would be interesting to analyze how the onset of the upturn correlates with the drive strength in the linear-response regime at a fixed distance away from the resonance. Moreover, since the cavity photons play a dual role of both driving the magnons and probing the system, they can be used to independently drive the ferromagnet near the magnon resonance, while using far-detuned resonator photons to probe the superconducting state. In this case, we expect the appearance of similar upturns whose quantitative dependence is determined by the functional form discussed in Eq.~\eqref{mzmain}.

{\it Acknowledgments.-} DC is supported in part by a grant from the Department of Energy (DE-SC0026112) under the Early Career Research Program, and DP acknowledges funding by the German Research Foundation (DFG) under Project-ID 442134789.

\bibliography{magnon_SC}
\bibliographystyle{apsrev4-2}

\clearpage

\begin{widetext}
\section{Supplementary material for ``Dynamical Control of Superconductivity in Superconductor–Ferromagnet Bilayers" }

\subsection{Driven magnon-polaritons}

In the rotating-wave approximation, the coupled photon-magnon system can be described by the Hamiltonian introduced in Eq.~\eqref{Hmp} of the main text. 
The driving of the resonator photon mode $\hat{a}$ by an external source can be modeled via a time-dependent Hamiltonian 
$H_\text{d}(t) = V_d \exp(-i\omega_d t) \hat{a}^\dagger + \text{h.c.}$, where we assume a monochromatic drive for simplicity.  As noted in the main text, the external drive generates a finite expectation value for the magnon operators via the magnon-photon hybridization, leading to a finite value for $m_z(t) \sim 
m_x/\sqrt{2S}\braket{\hat{b}(t) + \hat{b}^\dagger(t)}$, where $S$ is the macro-spin associated with the magnetization.

To determine this expectation value, it is convenient to revisit the magnon-polariton operators \cite{haug2009quantum}, i.e. the eigenmodes of Eq.\ \eqref{Hmp}, 
obtained via the unitary transformation 
\begin{align}
\hat{\alpha}_+ = u_+ \hat{a} + v_+ \hat{b}, \quad 
\hat{\alpha}_{-} = u_- \hat{a} + v_- \hat{b} , 
\end{align} 
with Hopfield (Boguliubov) coefficients 
\begin{align}
u_{\pm} &= \sqrt{\frac{\omega_\pm - \omega_m}{2\omega_\pm - \omega_r - \omega_m}}, \quad v_{\pm} = \pm i  \sqrt{\frac{\omega_\pm - \omega_r}{2\omega_\pm - \omega_r - \omega_m}} \ .  
    \end{align} 
The inverse transformation reads 
\begin{align}
\hat{a} &= r_+ \hat{\alpha}_+ + r_- \hat{\alpha}_-  \quad     && r_+= \frac{v_-}{u_+ v_- - u_- v_+ } \\ 
\hat{b} &= s_+ \hat{\alpha}_+ + s_- \hat{\alpha}_- 
 && r_-= \frac{v_+}{u_- v_+ - u_+ v_- }, 
\end{align}
and $s_\pm = r_\pm (u \leftrightarrow v)$.

In terms of the magnon-polariton operators, the driven Hamiltonian is 
\begin{align}
\label{Hmppolaritons}
H_\text{mp}(t) = H_\text{mp} + H_d(t) = \sum_{i = \pm}\omega_i \hat{\alpha}^\dagger_i \hat{\alpha}_i + V_d \exp(-i\omega_d t) (\overline{r}_+ \hat{\alpha}_+^\dagger + \overline{r}_- \hat{\alpha}_-^\dagger) + \text{h.c.}
\end{align}
Since the polariton operators in \eqref{Hmppolaritons} are decoupled from each other, one can conveniently determine the time evolution of $\hat{\alpha}_i$ in the Heisenberg picture, by making an ansatz $\hat{\alpha}_i(t) = \exp(-i\omega_i t) \left[\hat{\alpha}_i(0) + {c}_i(t)\right]$, where $c_i(t)$ is a time-dependent $c$-number. This allows to solve the Heisenberg equation of motion as 
\begin{align}
\label{polaritonevolution}
\hat{\alpha}_i(t) = \exp(-it\omega_i) \left[ \hat{\alpha}_i -i \int_0^t dt_1 \exp(i(\omega_i - \omega_d) t_1) \bar{r}_i V_d\right]  = \exp(-it\omega_i)\left(\hat{\alpha}_i - \bar{r}_i V_d \frac{\exp[i(\omega_i - \omega_d)t] - 1}{\omega_i - \omega_d}\right)
\end{align}
where we assume that the external drive  
is switched on for times $t > 0$. Given Eq.\ \eqref{polaritonevolution}, the magnon operators evolve as $\hat{b}(t) = \sum_i s_i \hat{\alpha}_i(t)$. As a result, the time-dependent magnon occupation can be found as 
\begin{align}
\braket{\hat{b}(t) + \hat{b}^\dagger(t)} = 2 \sum_i \text{Re} \left\{s_i \bar{r}_i V_d \frac{\exp(-i\omega_d t) - \exp(-i \omega_it)}{\omega_i - \omega_d} \right\}
\end{align}
Inserting the Hopfield coefficients $r_i,s_i$, we obtain 
\begin{align}
\label{bbdagger}
\braket{\hat{b}(t) + \hat{b}^\dagger(t)} =  \text{Im} \left\{ \frac{2gV_d}{\sqrt{4g^2 + (\omega_c - \omega_m)^2} } \frac{\exp(-i\omega_d t) - \exp(-i \omega_+t)}{\omega_+ - \omega_d} - \frac{2gV_d}{\sqrt{4g^2 + (\omega_c - \omega_m)^2} }\frac{\exp(-i\omega_d t) - \exp(-i \omega_-t)}{\omega_- - \omega_d} \right\} \ . 
\end{align}
Since the model in Eq.\ \eqref{Hmp} does not include any dissipation mechanism leading to a finite polariton life time, the expectation value in Eq.\ \eqref{bbdagger} diverges when $\omega_d \rightarrow \omega_\pm$. One can introduce dissipation by coupling the system to a bath and phrasing the time evolution in terms of a quantum master equation \cite{breuer2002theory}. Qualitatively, a finite polariton life time can also be accounted for by making the polariton frequencies complex, $\omega_i \rightarrow \omega_i - i\omega_i\kappa$, where $\omega_i\kappa$ is an effective decay rate, and $\kappa$ a dimensionless constant. For $t \gg (\omega_i \kappa)^{-1}$, the remaining time-dependence in Eq.\ \eqref{bbdagger} is governed by the external driving frequency $\omega_d$. When $\omega_d$ approaches a polariton frequency, e.g., $\omega_d \rightarrow \omega_+$, to the leading order in $\kappa$ one obtains, up to a phase shift
\begin{align}
\label{mzqm}
m_z (t) = \frac{m_x}{\sqrt{2S}}
\braket{\hat{b}(t) + \hat{b}^\dagger(t)} = \frac{\sqrt{2}g V_d m_x/\sqrt{S}}{\sqrt{4g^2 + (\omega_r - \omega_m)^2+(\omega_+ \kappa)^2} } \frac{1}{\sqrt{(\omega_d - \omega_+)^2+ (\omega_+ \kappa)^2}} \sin(\omega_d t), 
\end{align}
as used in the main text, Eq.\ \eqref{mzmain}.

\subsection{Semiclassical description of magnetization dynamics}

The main  features of the above result can also be derived semiclassically from the time evolution of the magnetization $\boldsymbol{m}$: We assume that the equilibrium orientation of $\vec{m}$ is set by a static magnetic field 
$\h_0 = h_x \e_x$, and model the drive as a dynamical field component  $\h_{\text{d}}(t)$, such that the total external field is $\boldsymbol{h}_\text{eff} = \h_0 + \h_{\text{d}}(t) $.  The resulting magnetization  is $\vec{m} = 
 m_x \vec{e}_x + \delta \m$,  where $|\delta \vec{m}| \ll |\m|$. 
 The dynamics of $\m$ can be captured via the LLG equation \cite{stancil2009spin, PhysRevLett.99.057003}
\begin{align}
\label{LLG}
\dot{\boldsymbol{m} }  = -\gamma (\boldsymbol{m} \times \boldsymbol{h}_\text{eff} ) + \frac{\kappa}{m_s}(\boldsymbol{m}  \times \dot{\boldsymbol{m}}), 
\end{align} 
where $m_s \simeq m_x$ is the saturation magnetization of the FM layer, $\gamma$ is the gyromagnetic ratio and $\kappa$ is an effective damping constant. For simplicity, we choose $\boldsymbol{h}_d(t) =  h_y \boldsymbol{e}_y \cos (\omega_d t)$. The solution for $m_z(t)$ then reads 
\begin{align}
m_z(t) = \text{Re} \left\{ \frac{ - i \omega_d h_y \gamma m_x }{\omega_d^2 - (\gamma h_x - i \omega \kappa)^2 } \exp(-i \omega_d t)\right \} 
\end{align}
If we replace the Larmor frequency $\gamma h_x$ by a phenomenological resonance frequency $\omega_+$ as in the quantum mechanical derivation, we find
\begin{align}
    m_z(t) \simeq  \frac{h_y \gamma m_x}{2 \sqrt{(\omega_d - \omega_+)^2 + ( \omega_+ \kappa)^2}} \sin (\omega_d t), 
\end{align}
where the last result holds up to phase shift and close to the resonance. This expression agrees with the quantum-mechanical one in Eq.\ \eqref{mzqm}
up to the different prefactor which encodes the magnon-photon hybridization. 

\subsection{Evaluation of the fermionic self-energy}

The effective BdG-Hamiltonian introduced in Eq.\ \eqref{Hamiltonian} is quadratic and can in principle be diagonalized exactly. However, to obtain a tractable description of the majority-spin physics, we obtain the self-energy for them perturbatively in first order in $\delta \hat{m}$, as graphically represented in Fig.\ \ref{magnondiag}.

\begin{figure}[H]
\centering
\includegraphics[width= 0.5\columnwidth]{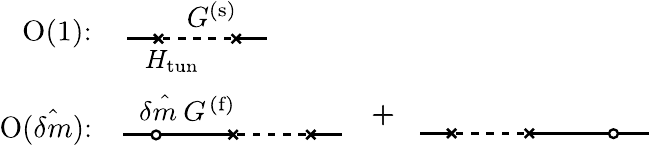}
\caption{Perturbative evaluation of the fermionic self-energy.  }
\label{magnondiag}
\end{figure}

In this figure, $G^{\text{(s)}}$ and  $G^{\text{(f)}}$ denote the matrix-valued propagators in the superconducting and ferromagnetic layers for $\omega = 0$, obtained by inverting the corresponding Hamiltonians, and a projector on the majority spins is applied on both sides of the diagram.

\subsection{Evaluation of $\Pi(\omega_r, T)$}

In the clean limit, the current-current correlation function $\Pi(\omega_r,T)$ for imaginary frequencies is evaluated as (see also \cite{PhysRevB.48.4219}): 
\begin{align} 
\label{Piapp}
&\Pi(i\omega , T) = - \braket{J_{\vec{u}} J_{\vec{u}}} (i\omega, T) =   \frac{1}{m^2}  \int_\k T\sum_{n}  k_{\vec{u}} (-k_{\vec{u}})     \left[G(\omega_+, \k) G(\omega_-,\k) + F(\omega_+,\k) F(\omega_-,\k) \right]   \ .  
\end{align} 
Here,$ \int_\k = \int\frac{d^2k}{(2\pi)^2}$ and   $\omega_{\pm} = \nu_n \pm \omega/2$. We have approximated the matrix element of the current operator as $J_{\vec{u}}(\k) = k_{\vec{u}}/m$, where $k_{\vec{u}}$ is the momentum component along the microwave vector potential.  $G$ and $F$ the normal and anomalous components of the spin-majority propagator in the Ferromagnet, defined as 
\begin{align} 
\label{bothprops}
G(i\omega,\k) = \frac{i\omega + \xi_\k}{\omega^2 + E_\k^2}, \quad   F(i\omega,\k) = \frac{\Delta_\k}{\omega^2 + E_\k^2}, 
\end{align} 
where $\xi_\k = k^2/2m - \mu$, $E_\k = \sqrt{\xi_\k^2 + |\Delta_\k|^2}$, and we have absorbed the in-plane magnetic field $h \simeq h_x$ in the chemical potential $\mu$. First, we perform the frequency summation in Eq.\ \eqref{Piapp}. When evaluating the pole residues, one should not simplify $n_F(E_\k + i\omega) = n_F(E_\k)$ despite $\omega$ being a bosonic Matsubara frequency, since this leads to the wrong result at $\omega = 0$ where the poles merge. After analytical continuation $i\omega \rightarrow \omega_r + i0^+$, we obtain 
\begin{align} 
\label{Piaftermat}
&\text{Re} \left[\Pi(\omega_r , T) \right]=   \frac{1}{2m^2}  \int_\k k_{\vec{u}}^2   \frac{\sinh{\frac{\omega_r}{T}}}{\omega_r \left(\cosh{\frac{E_\k}{T}} + \cosh{\frac{\omega_r}{T}}\right)} \ ,  
\end{align} 
which reduces to the textbook expression \cite{altland2010condensed} for the superfluid density when $\omega_r = 0$.  \\

\subsection{Frequency dependence of $\delta \omega_r(T)$}

In the evaluation in the main text,  $\delta \omega_r \sim \overline{\text{Re}\left[\Pi(\omega_r , T)\right]}$ is evaluated in the limit $\omega_r \rightarrow 0$. When $\omega_r > 0$, $\overline{\text{Re}\left[\Pi(\omega_r , T)\right]}$ does not vanish at $T = 0$, since a finite energy is supplied to the Bogoliubov quasiparticles. As can be seen from Fig.\ \ref{freqcomp}, which is obtained for $\Delta_2 = 0$, finite frequencies therefore flatten $\delta \omega_r(T)$.  A finite disorder has the same qualitative effect as a finite frequency, since it leads to a non-vanishing quasiparticle density of states at zero energy. Therefore, both finite frequency and disorder cannot be responsible for the temperature-dependent upturn observed in the experiment.

\begin{figure}
\centering
\includegraphics[width= 0.4\columnwidth]{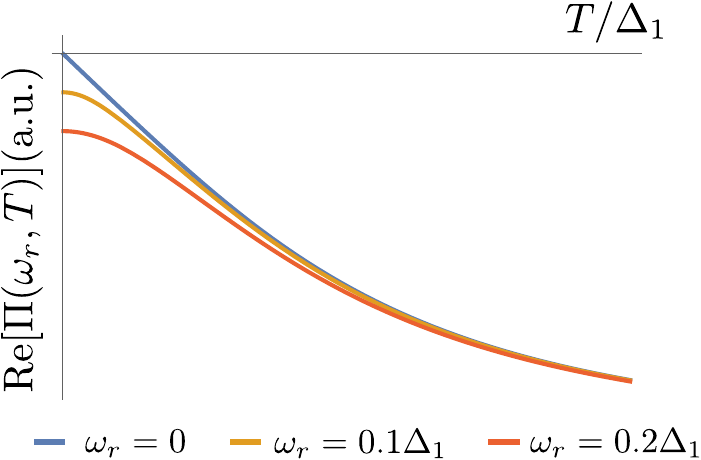}
\caption{$\text{Re}\left[\Pi(\omega_r,T) \right]$ from Eq.\ \eqref{Piaftermat} evaluated for a finite frequency $\omega_r$ and $\Delta_2 = 0$.    }
\label{freqcomp}
\end{figure}

\section{Effect of $T$-dependent scattering rate}

One aspect not taken into account so far is the finite lifetime of the fermions. Here, we include the lifetime in a heuristic way, by introducing a normal Matsubara self-energy (scattering rate)
\begin{align}
\label{GammaT}
    \Sigma(i\omega) = -i \text{sgn}(\omega) \Gamma(T), \quad \Gamma(T) \equiv  \Gamma_0 \exp(-\Delta^{*}/T) 
\end{align}

While the frequency dependence in Eq.\ \eqref{GammaT} has the same  form as in the standard Born-approximation result for disorder scattering, an additional key assumption  is the temperature dependence: We assume that the scattering phase space scales with the population of Bogoliubov quasiparticles and increases strongly (but smoothly) at temperatures exceeding $\Delta^{*}$. Such a functional dependence can arise from electron-electron scattering, or from scattering with a bosonic mode, e.g. the magnon-polariton. Furthermore, we assume that the magnitude of the rate $\Gamma_0$ scales with $\Delta^{*}$. 
This model can also be extended to the dynamic case by setting $\Gamma_0 \propto \Delta_2(t) = \Delta^{*} \sin(\omega_d t)$. 

To include the rate-effect in the evaluation of the frequency shift, we redefine the propagators of Eq.\ \eqref{bothprops} as 
\begin{align}
    G(i\omega, \k) &= \frac{i(\omega + \text{sgn}(\omega) \Gamma) + \xi_\k}{(\omega + \text{sgn}(\omega) \Gamma)^2 + E_\k^2} = \frac{u_\k^2}{i(\omega + \text{sgn}(\omega) \Gamma) - E_\k} + \frac{v_\k^2}{i(\omega + \text{sgn}(\omega) \Gamma) + E_\k}, \\
    F(i\omega,\k) &= \frac{\Delta_\k}{(\omega + \text{sgn}(\omega) \Gamma)^2 + E_\k^2} = - u_\k v_\k \left(\frac{1}{i(\omega + \text{sgn}(\omega) \Gamma) - E_\k} - \frac{1}{i(\omega + \text{sgn}(\omega) \Gamma) + E_\k}\right), 
\end{align}
where $u_\k, v_\k$ are the standard coherence factors \cite{mahan2000many}, and we suppress the $T$-dependence of $\Gamma$ for now. We evaluate $\Pi(i\omega,T) $ as defined in Eq.\ \eqref{Piapp} directly for vanishing external frequencies. For simplicity, we do not evaluate the vertex correction, which has been shown to lead to the qualitatively same results for $\Pi$ in the analysis of the disorder effect in \cite{boettcher2022new}.

The product of Green functions $GG + FF$ can be simplified by using $u_\k^2 + v_\k^2 = 1$. 
When taking the Matsubara sum, we now need to integrate along the real axis to avoid the discontinuity of $\text{sgn}(\omega) \Gamma$. 
We arrive at 
\begin{align}
\label{RePiscattering}
\Pi(\Gamma,T) \equiv \text{Re}[\Pi(\omega_r \rightarrow 0,T)]\bigg|_{\Gamma > 0} = \frac{1}{m^2} \int_{\k} k_{\boldsymbol{u}}^2 \int_{-\infty}^\infty \frac{d\epsilon}{\pi} \  n_F(\epsilon)
 \text{Im}\left(\frac{u_\k^2}{(\epsilon + i \Gamma - E_\k)^2} + \frac{v_\k^2}{(\epsilon + i \Gamma + E_\k)^2} \right) \ . 
\end{align}
Integrating \eqref{RePiscattering} by parts, we obtain
\begin{align}
\label{RePiscattering2}
\Pi(\Gamma,T) = -\frac{1}{m^2} \int_{\k} k_{\boldsymbol{u}}^2 \int_{-\infty}^\infty \frac{d\epsilon}{\pi} \  n_F^\prime(\epsilon) \frac{\Gamma}{(\epsilon - E_\k)^2 + \Gamma^2}, 
\end{align}
where we used that $n_F^\prime(\epsilon)$ is an even function. Evaluating the derivative, we obtain 
\begin{align}
\Pi(\Gamma,T) = \frac{1}{2m^2} \int_\k k_{\boldsymbol{u}}^2 \int_{-\infty}^\infty \frac{d\epsilon}{\pi} \frac{1}{T}\frac{1}{1 + \cosh\left(\frac{\epsilon}{T} \right) } \frac{\Gamma}{(\epsilon - E_\k)^2 + \Gamma^2}, 
\end{align}
which agrees with the previous result, Eq.\ \eqref{Piaftermat} in the limits $\omega_r \rightarrow 0, \Gamma \rightarrow 0$. One can see that temperature-dependence is smeared over a window of size $\Gamma$. 

In Fig.\ \ref{Gammafigure}, we show plots of $\Pi(\Gamma,T)$ for two values of $\Delta^{*}$. One can indeed observe an upturn: A finite scattering rate generally increases $|\Pi(\Gamma,T)|$. As temperatures are lowered, the scattering rate is switched off, which leads to a rapid suppression of $\Pi(\Gamma,T)$.

\begin{figure}
\centering
\includegraphics[width= \columnwidth]{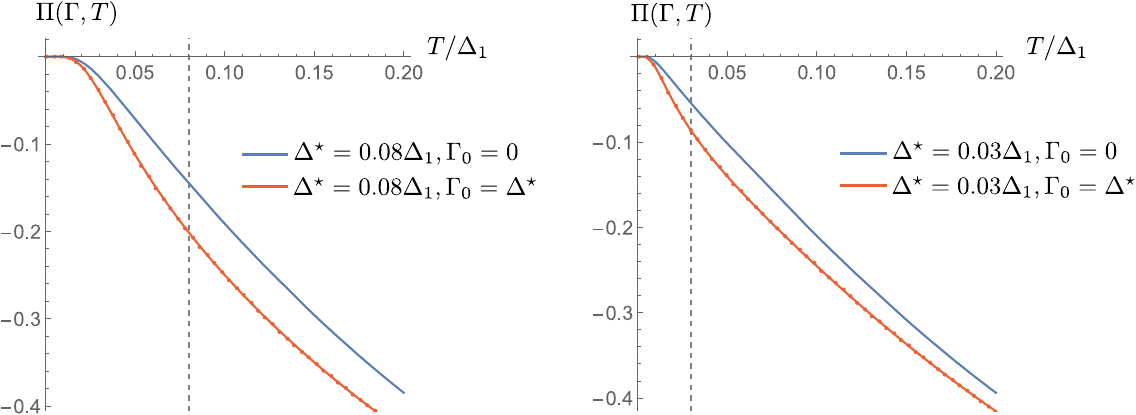}
\caption{$\Pi(\Gamma, T) \sim \delta \omega_r(T)$ for two values of $\Delta_2 = \Delta^{*}$, set to a constant time-independent value. The points are evaluated numerically.   }
\label{Gammafigure}
\end{figure}

\end{widetext}

\end{document}